\documentclass[submission,copyright,creativecommons]{eptcs}
\providecommand{\event}{Refine 2011} 
\usepackage{breakurl}             

\usepackage{tikz}
\usepackage{times,pstricks} 
\usepackage {pst-node}
\usepackage {latexsym}

\usepackage{hyperref}

\usepackage {stmaryrd}
\usepackage {amsmath}
\usepackage {amssymb}

\usepackage {floatflt}

\newcommand{\lend}[0]{\hspace{\fill}$\bullet$}

\newcommand {\eg}[0]{e.g.}
\newcommand {\ie}[0]{i.e.}

\newcommand {\pr} [1] {\ensuremath{{\sf #1}}}

\newcommand {\var} [1] {\ensuremath{{\mathbb #1}}}
\newcommand {\lvar} [1] {\ensuremath{\overline{\mathbb#1}}}

\newsavebox{\mybx}  
\newsavebox{\mybxx}  
\newsavebox{\mybxxx}  

\newlength{\dslen}
\newlength{\dslena}

\newsavebox{\mytab}

\newcommand {\paro}[4] {\ensuremath{
\settowidth{\dslen}{\hbox {{#2}  \hspace{1em} {#3}}}
{#1}\buildrel {{\Large #2} \hspace{1ex} {#3}} \over {\hbox to \dslen{\rightarrowfill}}{#4} }}

\newcommand {\mylogic} [3] {
\xymatrix@R=0pt@C=0pt@M=0pt{&{#1} \\
\ar@{-}[rr] &&\hspace{1.0ex}{#3}\hspace{1.0ex}\\
&{#2}}}

\def\rightharpfill{$\mathsurround=0pt - \mkern-6mu
\cleaders\hbox{$\mkern-2mu \mathord- \mkern-2mu$}\hfill
\mkern-6mu \mathord\rightharpoondown$}

\def\Rightarrowfill{$\mathsurround=0pt \mathord= \mkern-6mu
\cleaders\hbox{$\mkern-2mu \mathord= \mkern-2mu$}\hfill
\mkern-6mu \mathord\Rightarrow$}

\newcommand {\AroL}[4] {\ensuremath{
\settowidth{\dslen}{\hbox {$\langle{#2},{#4}\rangle$}}
{#1}\buildrel \langle{#2},{#4}\rangle \over {\hbox to \dslen{\Rightarrowfill}}{#3} }}

\newcommand {\Aroel}[4] {\ensuremath{
\settowidth{\dslen}{\hbox {${#2}$}}
{#1}\buildrel {#2} \over {\hbox to \dslen{\Rightarrowfill}}_{#4}{#3} }}

\newcommand {\aro} [3]
{\ensuremath{{#1}{\stackrel{#2}{\longrightarrow}}{#3} }}

\newcommand {\aroeltwo}[5] {\ensuremath{
\settowidth{\dslen}{\hbox {${#2}$}}
\settowidth{\dslena}{\hbox {${#4}$}}
{{#1}\buildrel {#2\hspace{-0.5ex}} \over {\hbox to \dslen{\rightarrowfill}}{\hspace{-0.5ex}#3}\buildrel {#4\hspace{-0.5ex}} \over {\hbox to \dslena{\rightarrowfill}}{\hspace{-0.5ex}#5} }}}

\newcommand {\Aro} [3]
{\ensuremath{{#1}{\stackrel{#2}{\Longrightarrow}}{#3} }}

\newcommand {\defeq}  {\ensuremath{ \hspace*{0.5em}
\stackrel{\mathrm{def}}{=} \hspace*{0.5em}} }

\newtheorem {mypdef}      {Definition}
\newtheorem {myplemma}     {Lemma} 
\newtheorem {myptheorem}  {Theorem}

\newtheorem {myex} {Example}

\newcommand {\sref} [1] {Section~\ref{sec:#1}}

\newcommand {\dref} [1] {Definition~\ref{def:#1}}

\newcommand {\lref} [1] {Lemma~\ref{lem:#1}}
\newcommand {\tref} [1] {Theorem~\ref{thm:#1}}
\newcommand {\fref} [1] {Figure~\ref{fig:#1}}

\newcommand {\mylend} {\noindent \ensuremath{\hfill\Box}}
\newcommand {\mydend} {\noindent \ensuremath{\hfill{\Box}}}

\newcommand {\aroelp}[4] {\ensuremath{
\settowidth{\dslen}{\hbox {${#2}\hspace{0.5em}$}}
{#1}\buildrel {#2} \over {\hbox to \dslen{\rightarrowfill}}_{#4}{#3} }}

\newcommand {\harp}[3] {\ensuremath{
\settowidth{\dslen}{\hbox {${#2}\hspace{0.5em}$}}
{#1}\buildrel {#2} \over {\hbox to \dslen{\rightharpfill}}{#3} }}

\newcommand {\saroelp}[4] {\ensuremath{\scriptstyle{
\settowidth{\dslen}{\hbox {${#2}$\hspace{-0.5em}}}
{#1}\buildrel {#2} \over {\hbox to \dslen{\rightarrowfill}}_{#4}{#3} }}}
\newcommand {\aroelps}[4] {\ensuremath{
\settowidth{\dslen}{\hbox {${#2}\hspace{-2.0em}$}}
{#1}\buildrel {#2} \over {\hbox to \dslen{\rightarrowfill}}_{#4}{#3} }}

\newcommand {\aroel}[3] {\ensuremath{
\settowidth{\dslen}{\hbox {${#2}$}}
{#1}\buildrel {#2\hspace{-0.5ex}} \over {\hbox to \dslen{\rightarrowfill}}{\hspace{-0.5ex}#3} }}
\newcommand {\arop}
[4]{\ensuremath{{#1}{\stackrel{#2}{\longrightarrow}_{#4}}{#3} }}

\newcommand {\Obsmath}[0]{\ensuremath{\mathbb O}}
\newcommand {\Emath}[0]{\ensuremath{\mathbb E}}

\usepackage {calc} 
\newcounter{hours}\newcounter{mins}

\title{Refinement for Probabilistic Systems with Nondeterminism}
\author{Steve Reeves
\institute{Department of Computer Science\\University of Waikato\\Hamilton\\New Zealand}
\email{stever@cs.waikato.ac.nz}
\and
David Streader
\institute{Department of Computer Science\\University of Waikato\\Hamilton\\New Zealand}
\email{\quad dstr@cs.waikato.ac.nz}
}

\def\titlerunning{Refinement for Probabilistic Systems with Nondeterminism}
\def\authorrunning{Steve Reeves \& David Streader}

\begin{document}
\maketitle


\begin{abstract}   
Before we combine actions and probabilities two very  obvious questions should be asked. Firstly, what does ``the probability of an action" mean?  Secondly, how does probability interact with nondeterminism?  Neither question has a single universally agreed upon answer but by considering these questions at the outset we build a novel and hopefully intuitive  probabilistic event-based formalism.

In previous work we have characterised refinement via the notion of testing. Basically, if one system passes all the tests that another system passes (and maybe more) we say the first system is a refinement of the second. This is, in our view, an important way of characterising refinement, via the question ``what sort of refinement should I be using?"

We use testing in this paper as the basis for our refinement. We develop tests for probabilistic systems by analogy with the tests developed for non-probabilistic systems. We make sure that our probabilistic tests, when performed on non-probabilistic automata, give us refinement relations which agree with for those non-probabilistic automata. We formalise this property as a vertical refinement.
\end{abstract}


\vspace{-.5cm}

\section{Introduction}\label{sec:intro}

Event-based models are frequently based on finite automata (FA, also called labelled transition systems) and probabilistic event-based systems are frequently based on FA where the transitions are also labelled by a probability as well as by an action.  Before we combine events and probabilities two very  obvious questions then arise. Firstly, what does ``the probability of an event" mean, or what does it mean for an event to ``behave in a probabilistic fashion"?  Secondly, how does probability interact with nondeterminism?  Neither question has a single universally agreed upon answer but by considering these questions at the outset we build a novel and hopefully intuitive  probabilistic event-based formalism.

Throughout we will be motivated by a wish to, in the end, develop a notion of refinement for probabilistic systems. In fact, refinement will be the starting point of our story here as well as the desired end point.


In previous work we have characterised refinement via the notion of testing. Basically, if one system passes all the tests that another system passes (and maybe more) we say the first system is a refinement of the second. This is, in our view, an important way of characterising refinement since the question ``what sort of refinement should I be using?" can be answered by saying ``you should be using the sort of refinement that is characterised by the sort of tests which characterise the contexts within which your system will find itself, \ie\ choose your refinement by looking at what contexts your systems will be used in."

Because this seems such a natural and useful answer, we use testing again in this paper as the basis for our refinement. We develop tests for probabilistic systems by analogy with the tests developed for non-probabilistic systems, all the while hoping to make sure that our probabilistic tests, when performed on non-probabilistic automata (and just noting whether a probability distribution is empty or not), give us refinement relations which agree with for those non-probabilistic automata: this gives us confidence that our new notions make sense. We formalise this property in \sref{vr}. 

The real test (!) in all this comes when we consider probabilistic automata which also contain nondeterminism. Again, we are guided by the wish that our probabilistic tests, when used on nondeterministic, non-probabilistic automata, give us a refinement ordering which agrees with that originally given for those automata when probability was not considered. We also find that the algebraic properties that characterise the non-probabilistic case carry over into our new domain.

We formalise a notion of refinement based upon probabilistic tests and then try to (re-)capture what nondeterminism means in this probabilistic setting.

We will first introduce transition systems as a semantic foundation for non-probabilistic automata and recap previous work on using testing to define refinement for such systems. 

It will turn out that part of the key to doing this for probabilistic systems is to be clear about two different philosophical bases for probability, so we next  review those. Another part of the key to this work will be a consideration of how nondeterminism is characterised, so we will go on to discuss that subsequently. This will finally suggest how we might adapt transition systems to allow consideration of probability, and we finally show how this adaptation can be used to also allow a treatment of nondeterministic probabilistic systems, all the while retaining our testing-based notion of refinement. 

We also show (via a selection) that expected properties hold for our refinement.

\section{Transition systems}

\begin{mypdef} \label{def:FA} 

Finite Automata (FA). Let $Act$ be a set of actions and let $Act^{\tau}$ be the same set along with the special action $\tau$, which represents actions interacting to form events. Let $N_{\sf A}$ be a finite set of nodes.

The finite automaton {\sf A} is given by the triple $(N_{\sf A}, S_{\sf A}, T_{\sf A})$ where
\begin{enumerate}
\item
$S_{\sf A} \subseteq  N_{\sf A}$ is a set of start nodes
\item
$T_{\sf A} \subseteq  \{ (n,{\sf a},m)| n,m \in N_{\sf A}\land {\sf a}\in Act^{\tau}\}$ shows the effect of each action.
\end{enumerate}
\end{mypdef}
We write \arop{x}{{\sf a}}{y}{\sf A} for $(x,{\sf a},y)\in T_{\sf A}$ and \aro{x}{{\sf a}}{y} where {\sf A} is obvious from context.
We write \aro{n}{\sf a}{}  for $\exists m.(n,{\sf a},m)\in T_{\sf A}$, 
and \aro{m}{\rho}{n}  for 
$$
\exists m_1\ldots m_i.\aro{m}{\rho_1}{m_1},\aro{m_1}{\rho_2}{m_2},\ldots \aro{m_i}{\rho_{i}}{n}
$$ 
and \aro{m}{\rho}{}  for 
$$
\exists m_1\ldots m_i,n.\aro{m}{\rho_1}{m_1},\aro{m_1}{\rho_2}{m_2},\ldots \aro{m_i}{\rho_{i}}{n}
$$ 
when $\rho = (\rho_1,...,\rho_i)$, a finite sequence of actions.

We write $\Aro{n}{}{m}$ for $\aro{n}{\tau^*}{m}$,  $\Aro{n}{\sf a}{m}$ for $\exists j, k. \Aro{n}{}{j} \land \aro{j}{\sf a}{k} \land \Aro{k}{}{m}$ and $\Aro{n}{\sf a}{}$ for $\exists j, k, m.\Aro{n}{}{j} \land \aro{j}{\sf a}{k} \land \Aro{k}{}{m}$.

$\Aro{m}{\rho}{}$  
and
$\Aro{m}{\rho}{n}$  
are defined similarly to the cases for \aro{}{}{}.

Where $\rho$ is a sequence of actions over $Act^{\tau}$ we write ${\rho_0}$ for $\rho$ with the $\tau$s removed.

\noindent The traces are $Tr({\sf A})\defeq \{ \rho\ |\  s \in S_{\sf A} \land \Aro{s}{\rho}{}\}$.

\noindent The complete traces\footnote{We deal with only acyclic automata and so we do not need to deal with infinite traces, though all the work of this paper can be extended to infinite traces and cyclic automata in the standard way \cite{Hen88}.} are $Tr^c({\sf A})\defeq \{ \rho\ |\ ( s \in S_{\sf A} \land \Aro{s}{\rho}{n} \land \pi(n)=\emptyset)$ where $\pi(n) \defeq \{m\ |\ \arop{n}{\sf x}{m}{\sf A}\}$.


We wish to model, using our automata, components that, like CSP processes,  can immediately be nondeterministic. But, unlike CSP, we wish hiding (abstraction) to distribute through choice (so $\tau$s are used only for unobservable actions or for events, and not pressed into service to encode nondeterministic choice between starting states). There is a subtle difference  between how external choice in CSP and choice in CCS behave with processes containing initial $\tau$ actions.  This has  been explained  either by regarding the choice operators as being different, see  \cite{LoN04} ``The unique choice operator of CCS, denoted by +, is a mixture between external and internal choices"  or by viewing CSP's use of $\tau$ actions to model a nondetermined  start state as different to CCS's use of $\tau$ actions \cite{ReS04}.   By  allowing automata to have a set of start states we both avoid having to distinguish external choice and CCS choice and allow hiding to distribute through choice \cite{ReS04}.

Also, choice can be defined (\cite{BaW90,WiN92}) between FAs with one start state each by gluing the two start states together to make a new single start state. Here, due to our generalisation, we glue together two sets of start states.

Let $S = \{s_1,s_2,\ldots ,s_n\}$ and $S' = \{s'_1,s'_2,\ldots ,s'_m\}$ be two sets of starting states and   then  define $\{S/ S\times S'\}$ to be the $n$ substitutions $\{s_i\in S |  s_i/ \{ (s_i,s'_1) ,\ldots , (s_i,s'_m) \}\}$  and define $\{S'/ S\times S'\}$ to be the $m$ substitutions $\{ s'_j\in S' | s'_j/ \{(s_1,s'_j) ,\ldots ,(s_n,s'_j) \}\}$.  


We define $\{SS'/ S\times S'\}$ to be the $n+m$ simultaneous substitutions 
$\{S/ S\times S'\}\cup \{S'/ S\times S'\}$.  The first $n$ substitutions replace each element of $\{s_1,s_2,\ldots ,s_n\}$ with a set of $m$ nodes and the last $m$ substitutions simultaneously replace each element of $\{s'_1,s'_2,\ldots ,s'_m\}$ with a set of $n$ nodes.
Consequently $\{S_{\sf A}S_{\pr B}/ S_{\sf A}\times S_{\sf B}\}$ will identify the two sets of nodes $S_{\sf A}$ and $S_{\sf B}$ as   $S_{\sf A}\{S_{\sf A}S_{\sf B}/ S_{\sf A}\times S_{\sf B}\}$ and $S_{\sf B}\{S_{\sf A}S_{\sf B}/ S_{\sf A}\times S_{\sf B}\}$ are both the $n\times m$ set of nodes $S_{\sf A}\times S_{\pr B}$. 

Since single states may now become sets of states under the substitution, we also have to define what it means to have sets of nodes in a transition:
$$
\aro{T}{\sf x}{T'} \defeq \{\aro{t}{\sf x}{t'} | t \in T, t' \in T' \}
$$


 \begin{mypdef} Process operators. Let {\sf A} be $(N_{\sf A},S_{\sf A},T_{\sf A})$ and let {\sf B} be $(N_ {\sf B}, S_ {\sf B}, T_ {\sf B})$.
 
\noindent Action Prefixing  ${\sf a}.{\sf B} = \defeq ( \{s\}\cup N_{\pr B}, \{s\}, \{\aro{s}{\sf a}{x} | x\in S_{\sf B}\}\cup T_{{\pr B}}  ) $  where $s$ is a new state.

 \noindent Internal choice
 ${\pr A}\sqcap  {\pr B} \defeq ( N_{{\pr A}}\cup N_{\pr B},S_{\sf A}\cup S_{\sf B} ,T_{{\pr A}}\cup T_{{\pr B}})$  
 
\noindent External choice is, informally, internal choice where start states are combined according to the substitutions above. Let $S_{{\pr A} \square {\pr B}}$ be $\bigcup((S_{\sf A}\cup S_{\sf B})\{S_{\sf A}S_{\pr B}/ S_{\sf A}\times S_{\sf B}\})$, \ie\ we combine start states as above. Then,

\noindent  External choice
${\pr A}\square {\pr B} \defeq ((N_{{\pr A}}\cup N_{\pr B}) \setminus (S_{\sf A}\cup S_{\sf B}) \cup S_{{\pr A} \square {\pr B}}, S_{{\pr A} \square {\pr B}}, (T_{{\pr A}}\cup T_{{\pr B}})\{S_{\sf A}S_{\pr B}/ S_{\sf A}\times S_{\sf B}\})$

\noindent Parallel composition: \label{def:Lop} 
${\sf A}\parallel_{P}{\sf B}\defeq (N_{{\sf A}\parallel_{P}{\sf B}},S_{{\sf A}\parallel_{P}{\sf B}},T_{{\sf A}\parallel_{P}{\sf B}})$ 
where $P \subseteq N_{\sf A} \cap N_{\sf B}$, $N_{{\sf A}\parallel_{P}{\sf B}} = N_{\sf A}\times  N_{\sf B}$, 
$S_{{\sf A}\parallel_{P}{\sf B}} = S_{\sf A} \times S_{\sf B}$ 
and $T_{{\sf A}\parallel_{P}{\sf B}}$  is defined by:

\noindent \hspace*{\fill}
\begin{tabular}{c}
\hspace{-1ex}$\arop{n}{{\sf x}}{l}{\sf A},\arop{m}{{\sf x}}{k}{\sf B},{\scriptstyle {\sf x}\in P}$
\hspace{-1ex} \\ \hline
$\arop{(n,m)}{\tau}{(l,k)}{{\sf A}\parallel_{P}{\sf B}}$
 \end{tabular} \hspace*{\fill}


\noindent 
\hspace*{\fill}
\begin{tabular}{c}
$\arop{n}{{\sf x}}{l}{\sf A},{\scriptstyle ({\sf x}\not\in P \land m\in N_{\sf B}})$ \\ \hline
$\arop{(n, m)}{{\sf x}}{(l, m)}{{\sf A}\parallel_{P}{\sf B}}$ 
\end{tabular}  
\hspace{\fill}
\begin{tabular}{c}
$\arop{n}{{\sf x}}{l}{\sf B},{\scriptstyle ({\sf x}\not\in P \land m\in N_{\sf A}})$ \\ \hline
$\arop{(m, n)}{{\sf x}}{(m, l)}{{\sf A}\parallel_{P}{\sf B}}$ 
\end{tabular} 
\hspace*{\fill}

%
%
%
\end{mypdef}

\begin{myex}
Let {\sf A} be 
$$
(\{s_1,s_2,t_1,t_2\},\{s_1,s_2\},\{\arop{s_1}{\sf a}{t_1}{\sf A},\arop{s_2}{\sf b}{t_2}{\sf A}\})
$$
and let {\sf B} be
$$
(\{s,s_2,t\},\{s\},\{\arop{s}{\sf c}{t}{\sf B}\})
$$
or, in diagram form,
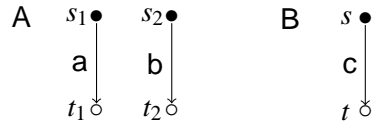
\begin{center}
\begin{tikzpicture} [inner sep=0pt]
\draw (0,3.25) node[](){\sf A};
  \draw (1,3.25) node[](s1){$\bullet$} 
  	    (.75,3.25) node[]{$s_1$}
             (1,2) node[](t1){$\circ$}
             (.75,2) node[]{$t_1$};
 \draw (2,3.25) node[](s2){$\bullet $} 
  	    (1.75,3.25) node[]{$s_2$}
             (2,2) node[](t2){$\circ$}
             (1.75,2) node[]{$t_2$};
  \draw [->] (s1)--(t1)node[left,pos=0.5,inner sep=3pt]{{\sf a}};
  \draw [->] (s2)--(t2)node[left,pos=0.5,inner sep=3pt]{{\sf b}};
  \end{tikzpicture} 
  \hspace{3 em}
  \begin{tikzpicture} [inner sep=0pt]
\draw (0,3.25) node[](){\sf B};
  \draw (1,3.25) node[](s){$\bullet $} 
  	    (.75,3.25) node[]{$s$}
             (1,2) node[](t){$\circ$}
             (.75,2) node[]{$t$};
 \draw [->] (s)--(t)node[left,pos=0.5,inner sep=3pt]{{\sf c}};
  \end{tikzpicture} 
\end{center}
Then ${\sf A} \sqcap {\sf B}$ is
$$
(\{s_1,s_2,s,t_1,t_2,t\},\{s_1,s_2,s\},\{\arop{s_1}{\sf a}{t_1}{\sf {\sf A} \sqcap {\sf B}},\arop{s_2}{\sf b}{t_2}{\sf {\sf A} \sqcap {\sf B}},\arop{s}{\sf c}{t}{\sf {\sf A} \sqcap {\sf B}}\})
$$
or, as a diagram,
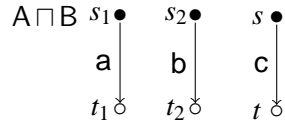
\begin{center}
\begin{tikzpicture} [inner sep=0pt]
\draw (0,3.25) node[](){${\sf A} \sqcap {\sf B}$};
  \draw (1,3.25) node[](s1){$\bullet $} 
  	    (.75,3.25) node[]{$s_1$}
             (1,2) node[](t1){$\circ$}
             (.75,2) node[]{$t_1$};
 \draw (2,3.25) node[](s2){$\bullet $} 
  	    (1.75,3.25) node[]{$s_2$}
             (2,2) node[](t2){$\circ$}
             (1.75,2) node[]{$t_2$};
  \draw [->] (s1)--(t1)node[left,pos=0.5,inner sep=3pt]{{\sf a}};
  \draw [->] (s2)--(t2)node[left,pos=0.5,inner sep=3pt]{{\sf b}};
  \end{tikzpicture} 
  \hspace{1 em}
  \begin{tikzpicture} [inner sep=0pt]
  \draw (1,3.25) node[](s){$\bullet $} 
  	    (.75,3.25) node[]{$s$}
             (1,2) node[](t){$\circ$}
             (.75,2) node[]{$t$};
 \draw [->] (s)--(t)node[left,pos=0.5,inner sep=3pt]{{\sf c}};
  \end{tikzpicture} 
\end{center}
Given that $S_{{\sf A} \square {\sf B}}$ is
$$
\bigcup \{s_1,s_2,s\}\{s_1/\{(s_1,s)\}, s_2/\{(s_2,s)\},s/\{(s_1,s),(s_2,s)\}\} = \{(s_1,s),(s_2,s)\}
$$ 
then  ${\sf A} \square {\sf B}$ is
$$
(\{t_2,t_3, (s_1,s),(s_2,s)\},\{(s_1,s),(s_2,s)\},
$$
$$
\{\arop{(s_1,s)}{\sf a}{t_1}{{\sf A} \square {\sf B}},\arop{(s_2,s)}{\sf b}{t_2}{{\sf A} \square {\sf B}},\arop{\{(s_1,s),(s_2,s)\}}{\sf c}{t}{{\sf A} \square {\sf B}}\})
$$
which is
$$
(\{t_2,t, (s_1,s),(s_2,s)\},\{(s_1,s),(s_2,s)\},
$$
$$
\{\arop{(s_1,s)}{\sf a}{t_1}{{\sf A} \square {\sf B}},\arop{(s_2,s)}{\sf b}{t_2}{{\sf A} \square {\sf B}},\arop{(s_1,s)}{\sf c}{t}{{\sf A} \square {\sf B}}\},\arop{(s_2,s)}{\sf c}{t}{{\sf A} \square {\sf B}}\})
$$
and as a diagram
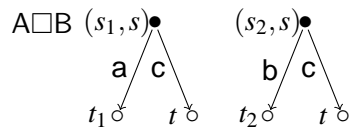
\begin{center}
\begin{tikzpicture} [inner sep=0pt]
\draw (-.5,3.25) node[](){${\sf A} \square {\sf B}$};
  \draw (1,3.25) node[](s1s){$\bullet $} 
  	    (.5,3.25) node[]{$(s_1,s)$}
             (.5,2) node[](t1){$\circ$}
             (.25,2) node[]{$t_1$};
 \draw (3,3.25) node[](s2s){$\bullet $} 
  	    (2.5,3.25) node[]{$(s_2,s)$}
             (2.5,2) node[](t2){$\circ$}
             (2.25,2) node[]{$t_2$};
  \draw [->] (s1s)--(t1)node[left,pos=0.5,inner sep=3pt]{{\sf a}};
  \draw [->] (s2s)--(t2)node[left,pos=0.5,inner sep=3pt]{{\sf b}};
  \draw  (1.5,2) node[](t31){$\circ$}
             (1.25,2) node[]{$t$};
   \draw (3.5,2) node[](t32){$\circ$}
             (3.25,2) node[]{$t$};
 \draw [->] (s1s)--(t31)node[left,pos=0.5,inner sep=3pt]{{\sf c}};
  \draw [->] (s2s)--(t32)node[left,pos=0.5,inner sep=3pt]{{\sf c}};
  \end{tikzpicture} 
\end{center}
Finally, ${\sf A}\parallel_{\{a\}}{\sf B}$ with (note that {\sf B}'s action is now {\sf a})
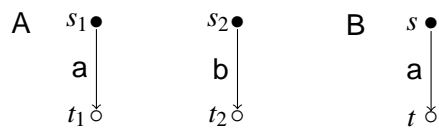
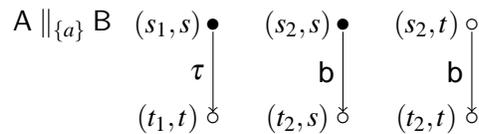
\begin{center}
\begin{tikzpicture} [inner sep=0pt]
\draw (0,3.25) node[](){\sf A};
  \draw (1,3.25) node[](s1){$\bullet $} 
  	    (.75,3.25) node[]{$s_1$}
             (1,2) node[](t1){$\circ$}
             (.75,2) node[]{$t_1$};
 \draw [->] (s1)--(t1)node[left,pos=0.5,inner sep=3pt]{{\sf a}};
  \end{tikzpicture} 
 \hspace{3 em}
  \begin{tikzpicture} [inner sep=0pt]
 \draw (1,3.25) node[](s2){$\bullet $} 
  	    (.75,3.25) node[]{$s_2$}
             (1,2) node[](t2){$\circ$}
             (.75,2) node[]{$t_2$};
  \draw [->] (s2)--(t2)node[left,pos=0.5,inner sep=3pt]{{\sf b}};
  \end{tikzpicture} 
  \hspace{3 em}
  \begin{tikzpicture} [inner sep=0pt]
\draw (0,3.25) node[](){\sf B};
  \draw (1,3.25) node[](s){$\bullet $} 
  	    (.75,3.25) node[]{$s$}
             (1,2) node[](t){$\circ$}
             (.75,2) node[]{$t$};
 \draw [->] (s)--(t)node[left,pos=0.5,inner sep=3pt]{{\sf a}};
  \end{tikzpicture} 
\end{center}
is
\begin{center}
\begin{tikzpicture} [inner sep=0pt]
\draw (0,3.25) node[](){${\sf A}\parallel_{\{a\}}{\sf B}$};
  \draw (2,3.25) node[](s1s){$\bullet $} 
  	    (1.4,3.25) node[]{$(s_1,s)$}
             (2,2) node[](t1t){$\circ$}
             (1.4,2) node[]{$(t_1,t)$};
   \draw [->] (s1s)--(t1t)node[left,pos=0.5,inner sep=3pt]{{$\tau$}};
   \end{tikzpicture} 
  \hspace{1 em}
  \begin{tikzpicture} [inner sep=0pt]
 \draw (2,3.25) node[](s2s){$\bullet $} 
  	    (1.4,3.25) node[]{$(s_2,s)$}
             (2,2) node[](t2s){$\circ$}
             (1.4,2) node[]{$(t_2,s)$};
  \draw [->] (s2s)--(t2s)node[left,pos=0.5,inner sep=3pt]{{\sf b}};
  \end{tikzpicture} 
  \hspace{1 em}
  \begin{tikzpicture} [inner sep=0pt]
  \draw (1,3.25) node[](s2t){$\circ$} 
  	    (.4,3.25) node[]{$(s_2,t)$}
             (1,2) node[](t2t3){$\circ$}
             (.4,2) node[]{$(t_2,t)$};
 \draw [->] (s)--(t)node[left,pos=0.5,inner sep=3pt]{{\sf b}};
  \end{tikzpicture} 
\end{center}
$\square$
\end{myex}

\section{Testing semantics}\label{sec:cbt}

The definitions in this section are taken from \cite{ReS09} where they have been applied to both state-based and event-based models.

One of our tests, of a process {\sf E}, taken from a set of processes $\Emath$, consists of placing {\sf E} in some context {\sf X} taken from a set of possible contexts $\Xi$. {\sf E} in context {\sf X} is  written  $[{\sf E}]_{\sf X}$. We then observe the resulting system.  Each observation made is taken from a set of possible observations $\Obsmath$.

We turn first to our general definition of testing semantics for nondeterministic processes and contexts. In this setting a test may return (nondeterministically) one observation from a set of possible observations. 

A specification is interpreted as a \emph{contract} consisting of the \emph{assumption} that the process will be placed only in one of the specified contexts $\Xi$ and a \emph{guarantee} that the observation of its behaviour will be one of the observations defined by the mapping $O: \Emath \rightarrow\Xi\rightarrow \wp \Obsmath$.  The mapping $O$ defines what can be observed  for all processes in any of the assumed contexts. Hence for any fixed $\Xi$ and $O$ we have a definition of the semantics and the refinement of processes.

\begin{mypdef}\label{def:ref} 

Let $\Xi$ be a set of contexts each of which the processes ${\sf A}, {\sf C} \in \Emath$ can communicate privately with, and let $O: \Emath \rightarrow\Xi\rightarrow \wp \Obsmath$ be a function which returns a set of observations, \ie\ a subset of $\Obsmath$. Then, the relational semantics of a process ${\sf A}$ is a subset of $\Xi \times \Obsmath$.
$$
 \llbracket {\sf A}\rrbracket_{\Xi,O} \defeq  \{(x,o) | x \in \Xi \wedge  o\in O([{\sf  A}]_x)\}
$$
and refinement is given by
 $$
{\sf A} \sqsubseteq_{\Xi, O} {\sf  C} \defeq   \llbracket {\sf C}\rrbracket_{\Xi,O} \subseteq  \llbracket {\sf A}\rrbracket_{\Xi,O}
$$
and equality is
$$
{\sf A} =_{\Xi, O} {\sf  C} \defeq   \llbracket {\sf C}\rrbracket_{\Xi,O} = \llbracket {\sf A}\rrbracket_{\Xi,O}
$$
\mylend
\end{mypdef} 

Given a rich enough class of tests  the use of nondeterministic tests is redundant, as what can be observed using a nondeterministic test will be the union of what can be observed using a set of deterministic tests. Hence nondeterministic tests add no further information and will be ignored. 


For all the processes considered in this paper, placing a process {\sf A} in a context {\sf X}, \ie\ ${[\sf  A}]_{\sf X}$, will mean executing process {\sf A} in parallel with {\sf X}, \ie\ ${\sf A}\parallel_N {\sf X}$ (where $N$ is some set of actions over which the context and process communicate, \ie\ synchronize) and the observation function $O$ is either the trace function $Tr$ (if only safety properties are of interest) or (if liveness properties are of interest) the complete trace function $Tr^c$.

\begin{mypdef}
Let   $\Xi_{FA}$ be FA  and  let  $\sqsubseteq_{FA} be \sqsubseteq_{\Xi_{FA}, Tr^c}$.
\mylend
\end{mypdef}

\begin{myptheorem} Refinement distributes through parallel composition: Let ${\sf X,Y,P,Q}\in{FA}$

\begin{center}\begin{tabular}{c}
${\sf X}\sqsubseteq_{FA}{\sf Y} , {\sf P}\sqsubseteq_{FA}{\sf Q}$ \\ \hline
${\sf X}\parallel_N {\sf P} \sqsubseteq_{FA} {\sf Y}\parallel_N{\sf Q}$
\end{tabular}\end{center}
\mylend
\end{myptheorem}

\section{Probabilities---Two Interpretations}\label{sec:four}

There are two (main) interpretations of probability, the \emph{frequentist}  and the \emph{Bayesian}. 

\begin{description}

\item[\it The frequentists'] definition sees probability as the long-run expected frequency of occurrence. The probability of event $A$ happening, where $n$ is the number of times event $A$ occurs in $N$ opportunities, is $P(A) = n/N$. 

\item[\it The Bayesians'] view of probability is related to degree of belief or state of knowledge. It is a measure of the plausibility of an event given incomplete knowledge. The Bayesian probabilist specifies some  given or assumed prior probabilities, which are then used in the computation of other probabilities. That is to say, anything that is nondeterministic or unknown must either be assigned some probability or have its probability computed from other, more primitive, known probabilities. Bayesian statisticians have developed several ``objective" methods for specifying prior probabilities. 
\end{description}

The frequentists' view is based upon repeatedly  performing the same  test many times  and, where the behaviour of the item under test is nondeterministic, aggregating the results of all the tests. Extending an event-based testing semantics  to record not just the set of possible observations but the probability with which they occur is a simple uniform way to extend event-based testing semantics to event-based  probabilistic testing semantics.  This can be further generalised  by representing both the process under test and the test process itself with  probabilistic automata. 

The Bayesian view fits well with Hoare's comment on nondeterminism \cite[p81]{Hoa85}:
\begin{quote}
``{\it There is nothing mysterious about this kind of nondeterminism: it arises from a deliberate decision to ignore the factors which influence the selection}''
\end{quote}

So, nondeterminism  in a process is merely a case of not having analysed it enough to quantify it, \ie\ attach to it some probabilities. Nondeterministic choice is probabilistic choice with unknown probabilities.  Surprisingly, this is not how testing semantics have been defined in the literature. 

As probabilities quantify (\ie\ attach a number to, or make quantitative) nondeterministic behaviour, it is clearly crucial when modelling some real process to distinguish between the behaviour of the process being deterministic and the behaviour being nondeterministic. Similarly when the process is observed interacting in some context it is crucial to distinguish the  nondeterminism of the process from the nondeterminism of the context.


Give a coin to a frequentist statistician and they experiment by flipping the coin a large number of times  noting down the number of times they observe heads being uppermost and the number of times they observe tails.  From this experiment they can compute the probability. 

An important point to note is that, to the frequentist, probabilities define how  likely  it is that an action is executed, or equivalently how likely it is that the execution ends in a particular state. The probability of an event occurring when the event  cannot be executed must be zero.  

The Bayesian statistician,  given a coin, knows that the only observations are heads and tails, and has no further information.  The skill of the Bayesian statistician is to assign a prior probability based on understanding the world that agrees with the frequentist. It becomes very important when we try to add probabilities to event-based processes that we either follow the frequentist and perform experiments (tests) or follow the Bayesian statistician and think clearly about the behaviour in the world of what we are modelling. 




\section{Probabilistic Finite Automata}

\subsection{Probability}

We introduce probabilities on choice by attaching probabilities to the start states of a process. There are two things to notice here: as in the non-probabilistic case with FAs, we represent nondeterminism on the initial state of a process by allowing the process to start in one of a \emph{set} of states; and we generalise this idea to represent the \emph{probability} of starting in some state of a process by attaching probabilities to each of its start states so that we can see what the probability of each possible start state being actually chosen for some particular execution of the process.

The first of these points is inherited from work \cite{ReS04a} which seeks to remove the need to use unobservable actions to also ``encode" or represent nondeterminism in a process by assuming the process makes an unobserved transition to its ``real" starting state (which may be one of many) from some single ``dummy" formal starting state. (And, of course, this is just a case of using the usual ``set of states" model uniformly for start states as well as all other states, which is something we are all familiar with from the ``classic" algorithm that constructs a deterministic finite-state automaton from a nondeterministic one.) Such unobserved actions can then be used exclusively to denote (synchronisation between) events. This idea is, in the second point above,  carried over into the probabilistic realm so that initial probabilistic choice is replaced by a probability distribution over the possible starting states. 

\begin{figure}[tbp]
\begin{center}
\begin{tikzpicture} [inner sep=0pt]
  \draw (1,3.25) node[](s){$\bullet $} 
  	    (.75,3.25) node[]{$s_1$}
	    (1.25,3.25) node[]{$p$}
             (1,2.5) node[] {$Q_1$}
             (1.65,2) node[](xr){} 
           (0.35,2) node[](xl){};
  \draw [-] (s)-- (xl) ;
  \draw [-] (s)-- (xr);
  \draw [-] (xl)-- (xr);
  \end{tikzpicture} 
  \hspace{0 em}
  \begin{tikzpicture} [inner sep=0pt]
  \draw (1,3.25) node[](s){$\bullet $}
             (.75,3.25) node[]{$s_2$}
	    (1.5,3.25) node[]{$1-p$}
             (1,2.5) node[] {$Q_2$}
             (1.65,2) node[](xr){} 
           (0.35,2) node[](xl){};
  \draw [-] (s)-- (xl) ;
  \draw [-] (s)-- (xr);
  \draw [-] (xl)-- (xr);
  \end{tikzpicture} 
  \hspace{2 em}
  \begin{tikzpicture} [inner sep=0pt]
  \draw (1.75,4.25) node[](t){$\bullet $}
             (1,3.25) node[](s1){$\circ$} 
              (1.5,4.25) node[]{$t$}
  	    (.75,3.25) node[]{$s_1$}
             (1,2.5) node[] {$Q_1$}
             (1.65,2) node[](xr1){} 
             (0.35,2) node[](xl1){};
  \draw [->] (t)--(s1)node[left,pos=0.5,inner sep=3pt]{{\sf a}}node[right,pos=0.5,inner sep=3pt]{$p$};
  \draw [-] (s1)-- (xl1) ;
  \draw [-] (s1)-- (xr1);
  \draw [-] (xl1)-- (xr1);
  \draw (2.5,3.25) node[](s2){$\circ$}
             (2.25,3.25) node[]{$s_2$}
             (2.5,2.5) node[] {$Q_2$}
             (3.05,2) node[](xr2){} 
           (1.85,2) node[](xl2){};
  \draw [->] (t)--(s2)node[left,pos=0.5,inner sep=3pt]{{\sf a}}node[right,pos=0.5,inner sep=3pt]{$1-p$};
  \draw [-] (s2)-- (xl2) ;
  \draw [-] (s2)-- (xr2);
  \draw [-] (xl2)-- (xr2);
  \end{tikzpicture} 
 \hspace{2em}
  \begin{tikzpicture} [inner sep=0pt]
  \draw (1,4.25) node[](s0){$\bullet $} 
             (1,3.25) node[](s1){$\circ$} 
  	    (.75,3.25) node[]{$s_1$}
	    (.75,4.25) node[]{$t_1$}
	    (1.25,4.25) node[]{$p$}
             (1,2.5) node[] {$Q_1$}
             (1.65,2) node[](xr){} 
           (0.35,2) node[](xl){};
  \draw [->] (s0)--(s1) node[left,pos=0.5,inner sep=3pt]{{\sf a}};
  \draw [-] (s1)--(xl) ;
  \draw [-] (s1)--(xr);
  \draw [-] (xl)--(xr);
  \end{tikzpicture} 
  \hspace{0 em}
  \begin{tikzpicture} [inner sep=0pt]
  \draw (1,4.25) node[](s0){$\bullet $} 
             (1,3.25) node[](s1){$\circ$} 
             (.75,3.25) node[]{$s_2$}
              (.75,4.25) node[]{$t_2$}
	    (1.5,4.25) node[]{$1-p$}
             (1,2.5) node[] {$Q_2$}
             (1.65,2) node[](xr){} 
           (0.35,2) node[](xl){};
  \draw [->] (s0)--(s1)node[left,pos=0.5,inner sep=3pt]{{\sf a}};
  \draw [-] (s1)-- (xl) ;
  \draw [-] (s1)-- (xr);
  \draw [-] (xl)-- (xr);
  \end{tikzpicture} 
 \caption{Probabilities on starting states}
\label{fig:PisQ1+pQ2}
\end{center}
\end{figure}
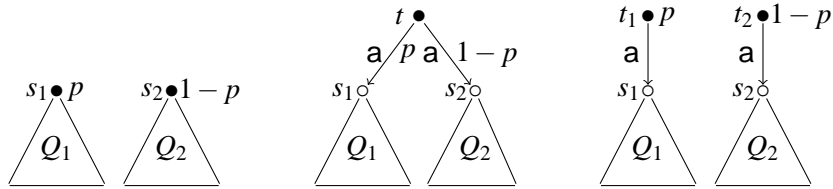

\begin{figure}[tbp]
\begin{center}
\begin{tikzpicture} [inner sep=0pt]
  \draw (1,3.25) node[](s){$\bullet $} 
  	    (.75,3.25) node[]{$s_1$}
	    (1.35,3.25) node[]{$p.q$}
             (1,2.5) node[] {$Q_1$}
             (1.65,2) node[](xr){} 
           (0.35,2) node[](xl){};
  \draw [-] (s)-- (xl) ;
  \draw [-] (s)-- (xr);
  \draw [-] (xl)-- (xr);
  \end{tikzpicture} 
  \hspace{0 em}
  \begin{tikzpicture} [inner sep=0pt]
  \draw (1,3.25) node[](s){$\bullet $}
             (.75,3.25) node[]{$s_2$}
	    (1.75,3.25) node[]{$(1-p).q$}
             (1,2.5) node[] {$Q_2$}
             (1.65,2) node[](xr){} 
           (0.35,2) node[](xl){};
  \draw [-] (s)-- (xl) ;
  \draw [-] (s)-- (xr);
  \draw [-] (xl)-- (xr);
  \end{tikzpicture} 
 \hspace{0em}
  \begin{tikzpicture} [inner sep=0pt]
  \draw  (1,3.25) node[](s1){$\bullet $} 
  	    (.75,3.25) node[]{$t$}
	    (1.45,3.25) node[]{$1-q$}
             (1,2.5) node[] {$P$}
             (1.65,2) node[](xr){} 
           (0.35,2) node[](xl){};
  \draw [-] (s1)--(xl) ;
  \draw [-] (s1)--(xr);
  \draw [-] (xl)--(xr);
  \end{tikzpicture} 
 \caption{More general probabilistic combination}
\label{fig:Q1+pQ2+qP}
\end{center}
\end{figure}
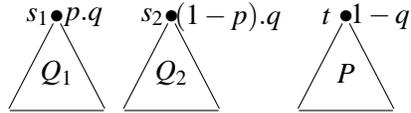

So, if $P$ is the process that starts with a choice between $Q_1$ and $Q_2$, which have (single, for this illustration) starting states $s_1$ and $s_2$ respectively, with probabilities $p$ of starting in state $s_1$ and $1-p$ of starting in state $s_2$ then we might picture $P$ as in the left of \fref{PisQ1+pQ2}.  We might represent the picture by saying $S(P) = \{s_1 \mapsto p, s_2 \mapsto 1-p\}$, where $S$ is a probability distribution function over start states of $P$.

Further, if we now form the process $a.P$ (\ie\ the event $a$ happens then the process $P$ happens) then we might picture this as in the middle of \fref{PisQ1+pQ2}, and here notice how the probabilities have migrated to the occurrences of event $a$. This picture suggests that transitions now represent the effect of an action on an initial state moving the system, according to some probability distribution, to the next state, when it synchronizes with the same action in some other process, \ie\ when the two actions combine to form an event which takes place with the indicated probability. 

So in $a.P$, the action $a$ has the potential to move us from state $t$ to state $s_1$ with probability $p$ and to $s_2$ with probability $1-p$ when synchronized to form an event which actually does take place with the indicated probabilities. We formalise all this by saying that the transitions of $a.P$ include $\{\aro{t}{\sf a}{d}\ |\  d(s_1) = p \land d(s_2) = 1-p\}$. An alternative picture might be as shown in the right of \fref{PisQ1+pQ2}, and here notice how the probabilities on the new start states for the new process $a.P$ have migrated from the old start states of $P$ and we have $S(a.P) = \{t_1 \mapsto p, t_2 \mapsto 1-p\}$. This picture might be considered a useful, though perhaps more unusual, alternative way of thinking of our system in the previous picture.  

Note that the original form of transitions as in FAs can be recovered by using the domain of the probability distribution function to tell us what the relevant post-states are.


As processes are combined together, the probabilities for the various component start states are combined to form the probabilities for the start states of the combination. As an example, see \fref{Q1+pQ2+qP}, which shows what the resultant start-state probabilities are for $(Q_1 +_p Q_2) +_q P$, where $s_1, s_2$ and $t$ are the start states for $Q_1, Q_2$ and $P$ respectively.

\subsection{Probability and nondeterminism}\label{sec:Prob&nondet}

From statistics, the \emph{law of large numbers} tells us that nondeterministic behaviour is the same as probabilistic behaviour where the probabilistic behaviour is unknown but can be found by repeating the right experiment a large number of times.

In process algebras $\tau$ actions indicate hidden, unobservable, uncontrollable actions or events (a special case being when two processes synchronize on some actions, which we consider to be private and uncontrollable).  Remember Hoare's comment that we cited in \sref{four}.  We have said above that we view this as agreeing with the Bayesian idea that probability  indicates a lack of information.

As probabilities refer to frequencies of executable behaviour, \ie\ the probability of an event occurring, they naturally occur on $\tau$ actions.  The intuitive relationship between nondeterminism and probability is widely held. For example,
\begin{quote}\emph{"nondeterminism represents possible choices that can be resolved
in a wholly unpredictable way. With probabilistic constructs the resolution
becomes predictable up to a point, in that it is quantified" }\cite{DGHMZ07}
\end{quote}

We can view this as saying that  probabilistic processes contain more information than nondeterministic processes but less than deterministic processes. Consequently what can be observed in any single observation of a probabilistic process is the same as what can be observed of the underlying non-probabilistic process. But by aggregating the observations of a large number of executions we can compute a probability distribution or verify a previously computed probability distribution.

As $\tau$ events are built by composing two actions that are observable (via parallel composition, \ie\ synchronization) it would be useful to find  some way to compute the probability of the executable $\tau$ event from the prior  ``probabilities" of their observable parts. This we do below in \dref{ops}.

The addition of probabilities to \emph{observable} actions where there is \emph{no} nondeterminism has proven both hard to interpret and hard to formalise, especially when we want to ensure that the models have desirable properties. One reason, in our opinion,  that this has turned out to be so hard to do is that the probabilities on the observable actions need, obviously, to define the behaviour of the processes not just in one context but in \emph{all} contexts.\footnote{We go no further with this point in this paper, but note that, in the non-probabilistic setting, we have considered this previously in \cite{ReS10}.}

\subsection{Nondeterminism}

We represent nondeterminism not by a separate set of operators but by allowing probabilities to be denoted not just by real numbers in the range 0 to 1 but also by real-valued terms (in that range) that contain variables or parameters. This introduces the idea of a starting-state distribution which is not completely determined or which has undetermined aspects, and hence allows us to represent nondeterminism with the same machinery that we introduce for probabilities.

This idea is motivated by the Bayesian view that the more we know about a mechanism, the more certain we can be about the probabilities attached to its behaviour: to talk of nondeterministic behaviour is merely to admit having more or less incomplete information about how something behaves, and this incomplete information can be represented by having parameters in the terms which denote probabilities. This also accords with Hoare's view that nondeterminism arises from ignoring or hiding (or, we would go further and say, being ignorant of) some aspects of a process. Further analysis of the mechanism would uncover (``unhide") more of the mechanism. This view dissolves nondeterminism; there is no such thing really, since it is just arises from not knowing (for whatever reason) enough about the actual distribution of probabilities amongst actions that might be taken when a choice is presented or confronted.

\subsection{Probabilistic testing semantics}\label{sec:cbpt}

For probabilistic tests all we need change is that the user records not just a set of observations but  a  probability distribution 
over a set of  observations,  hence $\Obsmath \defeq Act^*\rightarrow \mathbb{R}$.  

The relational semantics of process ${\sf A}$ when probability distributions are observed is a subset of $\Xi \times (Act^*\rightarrow \mathbb{R})$. If a process is experimented upon (frequentist perspective) and the results noted then what is observed will be  a function $\Xi \rightarrow(Act^*\rightarrow \mathbb{R})$  and hence there is no nondeterminism and no possibility  of refinement.

But approaching automata from the Bayesian perspective, if we can define the processes and tests as prior ``probabilistic" automata then we might be able to use probabilistic parallel composition to compute the probabilistic relational semantics of the processes. From the Bayesian point of view, the probabilities on actions are prior probabilities that, until the action takes part in an event by being synchronized with another process along the same action, do not play any role. Obviously the probability of an unexecuted action is prior to the probability of an execution---in particular, not until we factor in the probability of the synchronizing action do we know (via their product) what the probability of the executed event (denoted by $\tau$) will be. So, it is the Bayesian ideas that allow us to make sense of attaching probabilities to something that has not yet happened, and which will only be a part of what happens.

\section{Formalising  probabilistic automata}\label{sec:formprobaut}

In this section we will formalise the discussion in \sref{Prob&nondet} and see that automata that contain both probabilistic and nondeterministic choice are called \emph{partially probabilistic} introduced as  parameterised probabilistic finite automata (PPFA). Here we take what we see as the standard statistical approach and model nondeterministic choice as probabilistic choice with unknown probability. So our probabilities are no longer only real numbers but may also be real-valued terms (parameterised terms, hence the name) that may contain variables, the unknown probabilities. Automata where nondeterminism has been completely replaced by probabilistic choice are \emph{deterministic} probabilistic finite automata (DPFA).

\begin{mypdef} \label{def:PPFA} 
Parameterised Probabilistic Finite Automata  (PPFA ). Let $N_{\sf A}$ be a finite set of nodes. 
The parameterised probabilistic finite automaton {\sf A} is given by the triple $(N_{\sf A}, S_{\sf A}, T_{\sf A})$ where
\begin{enumerate}
\item
$S_{\sf A}$ is a ``starting  distribution", \ie\ a parameterised probability distribution such that $dom(S_{\sf A})\subseteq N_{\sf A}$, where $dom(S_A)$ are the starting states of {\sf A}
\item
$T_{\sf A} \subseteq \{ (n,{\sf a},d)| n \in N_{\sf A}\land {\sf a}\in Act^{\tau} \land d\in D_{\sf A}\}$, such that for  each $n\in N_{\sf A}$  and $a\in Act^{\tau}$ there exists no more than one element of $T_{\sf A}$ with first component $n$ and second component ${\sf a}$, and recall that nondeterminism is modelled by a parameter in the range of the probability distribution $d$. Finally, $D_A$ is a set of probability distributions over states.
\end{enumerate}

Deterministic Probabilistic Finite Automata  (DPFA) are PPFA with the restrictions that:
\begin{enumerate}
\item The ranges of all probability distributions are sets of real values, not sets of possibly parameterised terms, \ie\ the elements of the ranges contain no variables;
\item  $(n,{\sf a},d)\in T_{\sf A}$ implies ${\sf a}\in Act$ .
\end{enumerate}
\mydend
\end{mypdef}

Let the variables $\var{X},\var{Y}$ be taken from some set $Var$ and $\lvar{X}$ be a list of variables and $\psi_{\lvar{X}}$ be an instantiation of the variables in the list taken from the set of all such instantiations $\Psi_{\lvar{X}}$. We will write ${\sf A}(\lvar{X})$ for a PPFA containing variables $\lvar{X}$, but where not needed the list of variables will be dropped and we will write ${\sf A}$.  We interpret the variables in ${\sf A}(\lvar{X})$ as being \emph{globally bound} and take the usual $\alpha$-congruence of terms and identify PPFA that differ only by the names of variables used. Similarly we assume $\alpha$-renaming to prevent  confusion and variable capture when composing PPFAs.




We write \arop{x}{{\sf a}}{y}{{\sf A},p} for $(x,{\sf a},d)\in T_{\sf A}\land d(y) = p$ and \arop{x}{{\sf a}}{y} {p} where {\sf A} is obvious from context. In addition when we want to talk about a ``complete" transition, \ie\ one that has its associated final state distribution,  we write \arop{x}{{\sf a}}{d}{{\sf A}} for $(x,{\sf a},d)\in T_{\sf A}$.
 
\begin{mypdef}

The probability of the computation following a path, a sequence of transitions starting from a start state $s$, is the product of the probability of its component transitions and the probability of starting in the start state $S_{\sf A}(s)$. Let $p$ be the path $\arop{s}{\rho_1}{m_1}{p_1},\arop{m_1}{\rho_2}{m_2}{p_2},\ldots \arop{m_{n-1}}{\rho_{n}}{m_n}{p_{n}}$. Then the probability that $p$ is executed is 
$$
d(p) \defeq  S_{\sf A}(s) \times p_1\times p_2\times\ldots p_{n}
$$

and we say that the path $p$ can be observed as trace $\rho =\rho_1,\rho_2,\ldots ,\rho_{n}$.
 
The probability  of observing a trace $\rho$ is the sum of the all probabilities of the computation following any path that can be observed as trace $\rho$:

 $$
 d(\rho) = \displaystyle\sum_{tr(p_i) = \rho}d(p_i)
$$

where $tr(p) \defeq \{\rho | p = \arop{s}{\rho_1}{m_1}{p_1},\arop{m_1}{\rho_2}{m_2}{p_2},\ldots \arop{m_{n-1}}{\rho_{n}}{m_n}{p_{n}}\}$.
 
\end{mypdef}

Writing $\arop{S_{\sf A}}{\rho}{}{p}$  informs us that $p$ is the probability of seeing the trace $\rho$ when starting in any of the start states in $dom(S_{\sf A})$ and following some appropriate path, \ie\ $d(\rho)= p$. $\arop{S_{\sf A}}{\rho}{n}{p}$ means that $p$ is the probability of seeing the trace $\rho$ when starting in any of the start states in $dom(S_{\sf A})$ and ending in state $n$.

\begin{mypdef}
The probability distribution over complete traces is 
$$
D^c({\sf A})\defeq \{ \rho \mapsto \displaystyle\sum_{q \in P} q\ |\  P = \{q\ |\ n \in N_{\sf A} \land \pi(n)=\emptyset \land \arop{S_{\sf A}}{\rho}{n}{q} \}\}
$$
\end{mypdef}

\begin{mypdef} Process operators\label{def:ops}
 
 \noindent Action Prefixing  
 ${\sf a}.{\sf B} \defeq$ 
 $( \{s_a\} \cup N_{\pr B},\{s_a \mapsto 1\} ,\{\aro{s_a}{\sf a}{S_{\sf B}}\}  \cup T_{\pr B})$ where $s_{\sf a}$ is a new state

 \noindent Internal choice
 ${\pr A}\sqcap  {\pr B} \defeq ( N_{{\pr A}}\cup N_{\pr B},S_{\sf A} \sqcap S_{\sf B} ,T_{{\pr A}}\cup T_{{\pr B}} ) $  where

\noindent $(S_{\sf A} \sqcap S_{\sf B})(n) = \var{X} \times S_{\sf A}(n)$ if $n\in dom(S_{\sf A})$ else  $(1-\var{X})\times S_{\sf B}(n)$ if $n\in dom(S_{\sf B})$, where $\var{X}$ is a fresh parameter, and note that now $dom(S_{\sf A} \sqcap S_{\sf B}) = dom(S_{\sf A}) \cup dom(S_{\sf B})$.

  \noindent Probabilistic choice
 ${\pr A}\oplus_p  {\pr B} \defeq ( N_{{\pr A}}\cup N_{\pr B},S_{\sf A} \oplus_p  S_{\sf B} ,T_{{\pr A}}\cup T_{{\pr B}}) $   where $(S_{\sf A} \oplus_p  S_{\sf B}) (n) = p\times S_{\sf A}(n)$ if $n\in dom(S_{\sf A})$ else $ (1-p)\times S_{\sf B}(n)$ if $n\in dom(S_{\sf B})$, and note that now $dom(S_{\sf A} \oplus_p S_{\sf B}) = dom(S_{\sf A}) \cup dom(S_{\sf B})$. We note immediately from this that internal choice is probabilistic choice with unknown probability between the two choices.

 \noindent  External choice
${\pr A}\square {\pr B} \defeq ( N_{\pr A} \cup N_{\pr B} \setminus (dom(S_{\pr A}) \cup dom(S_{\pr B})) \cup dom(S_{{\pr A}\square {\pr B}}) ,S_{{\sf A} \square{\sf B}} ,T_{{\pr A}}\cup T_{{\pr B}}  \{\{S_{\pr A}S_{\pr B}/ S_{\pr A}\times S_{\pr B}\}\})$ 
where
$S_{{\sf A} \square{\sf B}}(n_{\sf A},n_{\sf B}) = S_{{\sf A}}(n_{\sf A})\times S_{{\sf B}}(n_{\sf B})$ 
and 
$\{\{S_{\pr A}S_{\pr B}/ S_{\pr A}\times S_{\pr B}\}\}$ now, of course, uses the \emph{domains} of the start state distributions in order to build the substitutions over start states.

\noindent Parallel composition: \label{def:PLop} 
$$
{\sf A}\parallel_{P}{\sf B}\defeq (N_{{\sf A}\parallel_{P}{\sf B}},S_{{\sf A}\parallel_{P}{\sf B}},T_{{\sf A}\parallel_{P}{\sf B}}) 
$$
$$
N_{{\sf A}\parallel_{P}{\sf B}} = N_{\sf A}\times  N_{\sf B}
$$
$$
S_{{\sf A}\parallel_{P}{\sf B}} (n_{\sf A},  n_{\sf B}) = S_{\sf A}(n_{\sf A})\times S_{\sf B}(n_{\sf B})\ if\ n_{\sf A} \in dom(S_{\sf A}) \land n_{\sf B} \in dom(S_{\sf B})
$$
and $T_{{\sf A}\parallel_{P}{\sf B}}$  is defined by:
$$
\begin{tabular}{c}
\hspace{-1ex}$\arop{n}{{\sf x}}{d_{\sf A}}{\sf A},\arop{m}{{\sf x}}{d_{\sf B}}{\sf B},{\scriptstyle {\sf x}\in P}$
\hspace{-1ex} \\ \hline
$\arop{(n,m)}{\tau}{d_{\sf A}\times d_{\sf B}}{({\sf A}\parallel_{P}{\sf B})}$
\end{tabular}
$$
$$
\begin{tabular}{c}
$\arop{n}{\sf x}{d_{\sf A}}{\sf A},{\scriptstyle ({\sf x}\not\in P \land m\in N_{\sf B})}$ \\ \hline
$\arop{(n, m)}{{\sf x}}{d_{\sf A}\times m}{({\sf A}\parallel_{P}{\sf B})}$ 
\end{tabular}
\hspace{3em}
\hspace{\fill}
\begin{tabular}{c}
$\arop{n}{{\sf x}}{d_{\sf B}}{{\sf B}},{\scriptstyle ({\sf x}\not\in P \land m\in N_{\sf A})}$ \\ \hline
$\arop{(m, n)}{{\sf x}}{m\times d_{\sf B}}{({\sf A}\parallel_{P}{\sf B})}$ 
\end{tabular}
$$
where 
$$
d_{\sf A}\times d_{\sf B} \defeq \{(x,y) \mapsto d_{\sf A}(x) . d_{\sf B}(y) | \arop{n}{\sf x}{x}{A} \land \arop{m}{\sf x}{y}{B}\}
$$
and
$$
d_{\sf A}\times m \defeq \{(x,m) \mapsto  d_{\sf A}(x) | \arop{n}{\sf x}{x}{\sf A}\}
$$  
and 
$$
m\times d_{\sf B} \defeq \{(m,y) \mapsto d_{\sf B}(y) | \arop{n}{\sf x}{y}{\sf B}\}
$$  
\end{mypdef}

\begin{myex}
Consider the PPFAs given by the expressions $a.(Q_1 +_p Q_2)$ and $a.Q_1 +_p a.Q_2$. Then, assuming the start states, states and transitions of $Q_1$ and $Q_2$ are given by $s_1$, $s_2$, $N_1$, $N_2$, $T_1$ and $T_2$ respectively, we have
$$
a.Q_1 +_p a.Q_2 = (\{t_1, t_2\} \cup N_1 \cup N_2, \{t_1 \mapsto p, t_2 \mapsto 1-p\}, 
$$
$$
\{\aro{t_1}{\sf a}{d_1}, \aro{t_2}{\sf a}{d_2} | d_1(s_1)=  d_2(s_2) = 1\} \cup T_1 \cup T_2)
$$

$$
a.(Q_1 +_p Q_2)= (\{t\} \cup N_1 \cup N_2, \{t \mapsto 1\},
$$
$$
 \{\aro{t}{\sf a}{d} | d(s_1)=  p, d(s_2) = 1-p\} \cup T_1 \cup T_2)
$$
\end{myex}
In fact, these PPFAs are indistinguishable by testing, so they are equal (they ``refine both ways") as far as our testing semantics goes. This result can be generalised so that probability distributions on transitions can always be ``migrated" to the starting state distribution.

\subsection{Testing of probabilistic processes}\label{sec:cbtpp}

Recall from \sref{cbt} that we said in the definition of our testing semantics for FA that we will use ${[\sf  A}]_{\sf X} = {\sf A}\parallel_N {\sf X}$ and $O_{FA}= Tr^c$. For probabilistic FAs we  need to use parallel composition from   \dref{PLop} (as defined for DPFA and PPFA). The observation of a single execution of a DPFA is still a trace but what can be ``observed" over many executions is  no longer simply a set of traces but, if we also record the frequency of occurrence of the traces, a probability distribution over the set of traces hence $O_{DPFA}=  D^c$.  We treat PPFA similarly and let   $\Xi_{PPFA} =PPFA$  and $O_{PPFA} =  D^c$  except that now the observed probability distributions may be parameterised.  

\begin{mypdef}\label{def:pref}

The relational semantics of an entity ${\sf A}(\lvar{X})$ is (where $\Psi_{\lvar X}$ is the set of instantiations for the parameters in $\lvar X$)
$$
 \llbracket {\sf A(\lvar{X})}\rrbracket_{\Xi_{PPFA}, D^c} \defeq  \{(x,o).x \in \Xi_{PPFA} \wedge  o\in \psi_{\lvar{X}}(D^c(([{\sf  A}(\lvar{X})]_x))) \wedge \psi_{\lvar{X}}\in \Psi_{\lvar{X}}\}
 $$
$$
{\sf A}(\lvar{X}) \sqsubseteq_{\Xi_{PPFA}, D^c} {\sf  C}(\lvar{Y}) \defeq   \llbracket{\sf C}(\lvar{Y})\rrbracket_{\Xi_{PPFA}, D^c} \subseteq \llbracket {\sf  A}(\lvar{X})\rrbracket_{\Xi_{PPFA}, D^c}
$$
$$
{\sf A}(\lvar{X}) =_{PPFA} {\sf  C}(\lvar{Y}) \defeq   \llbracket{\sf C}(\lvar{Y})\rrbracket_{\Xi_{PPFA}, D^c} = \llbracket {\sf  A}(\lvar{X})\rrbracket_{\Xi_{PPFA}, D^c}
$$

\end{mypdef}

Note here that we have  given the meaning of PPFAs as a relation from contexts (PPFAs) to  probability distributions:
$$
\llbracket {\sf A(\lvar{X})}\rrbracket_{\Xi_{PPFA},D^c} \subseteq \Xi_{PPFA} \times (Act^*\rightarrow {Real})
$$
by instantiating all the open distributions that might be observed to get plain probability distributions ``with no unknowns".

Let $\sqsubseteq_{PPFA} \defeq \sqsubseteq_{\Xi_{PPFA}, D^c}$.  That is, we  write $\sqsubseteq_{PPFA}$ for this  general definition of refinement.  When $\sqsubseteq_{PPFA}$ relates two  DPFA processes  it is of little interest, \ie\ there are no opportunities for refinement as there is no nondeterminism (though there are, perhaps, probabilities).

In \sref{vr} we will show refinement of PPFA is strongly related to refinement of an underlying FA.

\subsection{Simple results from the definitions}

\begin{myptheorem} Refinement distributes through parallel composition. Let {\sf X, Y, P} and {\sf Q} be arbitrary PPFAs and let $N \subseteq Act$. Then

\begin{center}\begin{tabular}{c}
${\sf X}\sqsubseteq_{PPFA}{\sf Y} , {\sf P}\sqsubseteq_{PPFA}{\sf Q}$ \\ \hline
${\sf X}\parallel_N {\sf P} \sqsubseteq_{PPFA} {\sf Y}\parallel_N{\sf Q}$
\end{tabular}\end{center}

\end{myptheorem}

\noindent For an arbitrary PPFA ${\sf P(\lvar Y)}$ we have the following theorems.

\begin{myptheorem} $\sqcap$ is idempotent. \label{thm:sqcap}
${\sf P}(\lvar{Y}) =_ {PPFA} {\sf P}(\lvar{Y})\sqcap {\sf P}(\lvar{Y})$

\end{myptheorem}
\noindent Proof: From \dref{PLop} it can be seen that the  graph of ${\sf P}(\lvar{Y})\sqcap {\sf P}(\lvar{Y})$ consists of two copies of the graph of ${\sf P}(\lvar{Y}) $ which ever copy is selected the behaviour is exactly that of ${\sf P}(\lvar{Y}) $. Hence he equality.

\begin{myptheorem} $\oplus_p$ is idempotent  
$ {\sf P}(\lvar{Y})  =_ {PPFA} {\sf P}(\lvar{Y}) \oplus_p {\sf P}(\lvar{Y})$

\end{myptheorem}
\noindent Proof:  Similar to  \tref{sqcap}.         

%
%

 \section{Relating finite automata  to parameterised probabilistic finite automata}\label{sec:vr}
 
We construct  $\llbracket\_\rrbracket_{PPFA}^{FA}$,  an embedding of FA into PPFA and a forgetful mapping from PPFA to FA, and then show that these mappings form a Galois connection between the refinement relations $\sqsubseteq_{PPFA} $ and $\sqsubseteq_{FA} $. 
 
The embedding  $\llbracket\_\rrbracket_{PPFA}^{FA}$  of FA in PPFA  will map  all nondeterministic choices in FA processes into probabilistic choice with unknown probabilities in the PPFA processes.
 
 \begin{mypdef}\label{def:sat} \label{def:imp} 
 
Semantic mappings $\llbracket\_\rrbracket_{PPFA}^{FA}$  and $vA^{FA}_{PPFA}$ between  finite automata ${\sf A}$ and  parameterised probabilistic finite automata  ${\sf Ap}$  are defined so that:
$$
\llbracket(N_{\sf A}, S_{\sf A}, T_{\sf A}) \rrbracket_{PPFA}^{FA} \defeq (N_{\sf Ap}, S_{\sf Ap}, T_{\sf Ap}) 
$$ 
where
$$
N_{\sf Ap} \defeq N_{\sf A}
$$
and 
$$
S_{\sf Ap} \defeq \{ (s,\var{X})\ |\   s\in S_{\sf A}  \land  \var{X}\ is\ fresh \land
(\displaystyle\Sigma_{n\in dom(S_{\sf Ap})} S_{\sf Ap}(n)) = 1\}
 $$
$$ 
 T_{\sf Ap}  = \{ (n,{\sf a},d)\ |\  d = \{m \mapsto v \ |\ \aro{n}{\sf a}{m} \land  v\ is\ fresh\}  \land
 (\displaystyle\Sigma_{m\in dom(d)} d(m)) = 1\}
$$
 
 \noindent The mapping $vA^{FA}_{PPFA}$ from {PPFA} in to {FA}  forgets all probability distributions: 
 
 $$
 vA^{FA}_{PPFA}(N_{\sf Ap}, s_{\sf Ap}, T_{\sf Ap}) = (N_{\sf A}, s_{\sf A}, T_{\sf A})
 $$ 
 where
$$
N_{\sf A} \defeq N_{\sf Ap}
$$  
and 
$$
S_{\sf A} \defeq  dom(S_{\sf Sp})
$$ 
and 
 $$
 T_{\sf A}  = \{ (n,{\sf a},m) |  \arop{n}{\sf a}{d}{{\sf Ap}} \land  m \in dom(d)\}
 $$
\mydend \end{mypdef}
  
The pair of mappings $(\llbracket\_\rrbracket_{PPFA}^{FA}, vA^{FA}_{PPFA})$ define a vertical refinement $\sqsubseteq^{FA}_{PPFA}$ as they are a Galois connection \cite{ReS10}. This is the content of \tref{GC}, but first some preliminary results.

\begin{myplemma}\label{lem:embed} For any FAs {\sf X} and {\sf Y}

$Tr^c({\sf X})\subseteq Tr^c({\sf Y}) \Rightarrow D^c(\llbracket{\sf X}\rrbracket_{PPFA}^{FA}) \subseteq D^c(\llbracket{\sf Y}\rrbracket_{PPFA}^{FA}) $

\end{myplemma}

Proof (Sketch)  The application of $\llbracket\_\rrbracket_{PPFA}^{FA}$ to a FA simply adds parameterised probabilities spanning any nondeterministic choice.  The set of all possible observation traces is $Tr^c({\sf X})$. 
This is also the set of all possible observation traces of $\llbracket{\sf X}\rrbracket_{PPFA}^{FA}$ but now what is ``observed" is not one trace but any probability distribution over  any subset of $O({\sf X})$ (we need to use subset as  when the probability of observing a trace is 0 it is no longer in the domain of the distribution). 

Hence $d\in D^c(\llbracket{\sf X}\rrbracket_{PPFA}^{FA}) \Leftrightarrow dom(d)\subseteq Tr^c({\sf X})$.
Consequently if $d\in D^c(\llbracket{\sf X}\rrbracket_{PPFA}^{FA}) $ then $dom(d)\subseteq Tr^c({\sf X})$ and since $Tr^c({\sf X})\subseteq Tr^c({\sf Y})$, from the assumption of the lemma, we further have $dom(d)\subseteq Tr^c({\sf Y})$. Then $d\in D^c(\llbracket{\sf Y}\rrbracket_{PPFA}^{FA}) $ follows from the argument above with {\sf Y} in place of {\sf X}. \lend
  
\begin{myptheorem}\label{thm:congpar} 

Let {\sf X} and {\sf Y} be FAs, and let $N \subseteq Act$. Then,
$$ 
\llbracket{\sf X}\parallel_N{\sf Y}\rrbracket_{PPFA}^{FA}
=
\llbracket{\sf X}\rrbracket_{PPFA}^{FA}\parallel_N\llbracket{\sf Y}\rrbracket_{PPFA}^{FA}
$$

\end{myptheorem}

\begin{myptheorem}\label{thm:congpar2}

Let  {\sf X} and {\sf Y} be PPFAs, and let $N \subseteq Act$. Then,
$$ 
vA^{FA}_{PPFA}({\sf X}\parallel_N{\sf Y})
=
vA^{FA}_{PPFA}({\sf X})\parallel_N vA^{FA}_{PPFA}({\sf Y})
$$

\end{myptheorem}

\begin{mypdef} Deterministic automata.

$Det_{FA} \defeq \{ {\sf P} | (\aro{n}{\sf a}{k} \wedge  \aro{n}{\sf a}{l} \Rightarrow k=l) \wedge |S_{\sf A}| =1\}$

$Det_{PPFA} \defeq \{ {\sf P} | (\arop{n}{\sf a}{k}{p} \wedge  \arop{n}{\sf a}{l}{q} \Rightarrow k=l \wedge p=q=1) \wedge |S_{\sf A}| =1\}$
\end{mypdef}

\begin{myplemma}\label{lem:suf} Results involving deterministic automata.

\begin{enumerate}
\item 
\begin{enumerate}
\item
$\{ {\sf X} \in Det_{FA}\ |\ \llbracket{\sf X}\rrbracket_{PPFA}^{FA}\} = Det_{PPFA}$ and
\item 
$\{ {\sf Y} \in Det_{PPFA}\ |\ vA^{FA}_{PPFA} ({\sf Y})\}= Det_{FA}$
\end{enumerate}

\item Let {\sf A} and {\sf C} be FAs. Then ${\sf A}\sqsubseteq_{FA}  {\sf C}\Leftrightarrow \forall_{x \in Det_{FA}}. Tr^c([{\sf  A}]_x) \supseteq Tr^c([{\sf  C}]_x)$

\item Let {\sf A} and {\sf C} be PPFAs. Then ${\sf A}\sqsubseteq_{PPFA}  {\sf C}\Leftrightarrow \forall_{x \in Det_{PPFA}}. D^c([{\sf  A}]_x) \supseteq D^c([{\sf  C}]_x)$
\end{enumerate}
\end{myplemma}

\noindent Proof (Sketch).

$1(a)$ and $1(b)$ follow from definitions.

Re 2: With non-probabilistic processes and tests, what can be observed when applying a nondeterministic test is the union of what can be observed when applying each element of the set of deterministic alternatives (where here we picture, as usual, a nondeterministic computation as a set of deterministic ones which covers all the possible choices)  and hence:
$$
{\sf A}\sqsubseteq_{FA}  {\sf C}\Leftrightarrow \forall_{x \in Det_{FA}}. Tr^c([{\sf  A}]_x) \supseteq Tr^c([{\sf  C}]_x)
$$

Re 3: With probabilistic processes and tests, what can be observed when applying a probabilistic test is the distribution, inferred  from the test, of what can be observed when applying the deterministic components that the probabilistic choice spans. Hence a  set of test processes for PPFA that is sufficient to establish refinement is the image after applying  $\llbracket{\_}\rrbracket_{PPFA}^{FA}$ to a sufficient  set of  FA processes, \ie\ since $Det_{FA}$ is sufficient for FA then $Det_{PPFA}$ is sufficient for PPFA, hence:
$$
{\sf A}\sqsubseteq_{PPFA}  {\sf C}\Leftrightarrow \forall_{x \in Det_{PPFA}}. D^c([{\sf  A}]_x) \supseteq D^c([{\sf  C}]_x)
$$

\begin{myptheorem}\label{thm:GC}
$$ 
\forall {\sf X}\in FA, {\sf Y}\in PPFA.  \llbracket{\sf X}\rrbracket_{PPFA}^{FA}\sqsubseteq_{PPFA}  {\sf Y}\Leftrightarrow {\sf X}\sqsubseteq_{FA}  vA^{FA}_{PPFA}({\sf Y})
$$
\end{myptheorem}

\noindent Proof:  (Sketch)

It is a well-known result (\eg\ \cite{Tay99}) that to prove a Galois connection it is sufficient to prove for arbitrary {\sf X}
$$
vA^{FA}_{PPFA}(\llbracket{\sf X}\rrbracket_{PPFA}^{FA}) \sqsubseteq_{FA}  id_{FA} {\sf X}
$$
and for arbitrary {\sf Y}
\[
\llbracket{vA^{FA}_{PPFA}(\sf Y)}\rrbracket_{PPFA}^{FA} \sqsubseteq_{PPFA}  id_{PPFA}{\sf Y}
\]
and in addition to prove both relations $\llbracket{\_}\rrbracket_{PPFA}^{FA}$ and $vA^{FA}_{PPFA}$ are monotone.

We can see directly from the definitions that   $\llbracket{\_}\rrbracket_{PPFA}^{FA}$ adds parameterised probabilities to any nondeterministic choice and $vA^{FA}_{PPFA}$ forgets this addition hence, for arbitrary {\sf X} :
$$
vA^{FA}_{PPFA}(\llbracket{\sf X}\rrbracket_{PPFA}^{FA}) =_{FA}  id_{FA}{\sf X}
$$
which gives our first inequality.

The effect of $\llbracket{vA^{FA}_{PPFA}{\sf Y}}\rrbracket_{PPFA}^{FA}$  is to first replace probabilistic choice with nondeterministic choice (by ignoring probabilities) and then reintroducing probabilities-with-parameters due to the nondeterminism and this can be refined, along with other possibilities,  back into its original value, which gives our second inequality.

Re:  show $\llbracket{\_}\rrbracket_{PPFA}^{FA}$ is monotone:   ${\sf A}\sqsubseteq_{FA}  {\sf C}\Rightarrow \llbracket{\sf A}\rrbracket_{PPFA}^{FA}\sqsubseteq_{PPFA}  \llbracket{\sf C}\rrbracket_{PPFA}^{FA}$

From \dref{ref} we have ${\sf A}\sqsubseteq_{FA}  {\sf C}\Leftrightarrow \forall_{x \in \Xi_{FA}}. Tr^c([{\sf  A}]_x) \supseteq Tr^c([{\sf  C}]_x)$ and as  $Det_{FA}\subseteq \Xi_{FA}$ we also have

${\sf A}\sqsubseteq_{FA}  {\sf C}\Leftrightarrow \forall_{x \in Det_{FA}}. Tr^c([{\sf  A}]_x) \supseteq Tr^c([{\sf  C}]_x)$ \hspace{\fill}(1)

%

From \lref{embed} we then have 

${\sf A}\sqsubseteq_{FA}  {\sf C}\Rightarrow  \forall_{x \in Det_{FA}}. D^c(\llbracket[{\sf A}]_{x}\rrbracket_{PPFA}^{FA}) \supseteq D^c(\llbracket [{\sf C}]_{x}\rrbracket_{PPFA}^{FA})$.

Then,

$ \forall_{x \in Det_{FA}}. D^c(\llbracket[{\sf A}]_{x}\rrbracket_{PPFA}^{FA}) \supseteq D^c(\llbracket [{\sf C}]_{x}\rrbracket_{PPFA}^{FA})$

$ \forall_{x \in Det_{FA}}. D^c([\llbracket{\sf A}\rrbracket_{PPFA}^{FA}]_{\llbracket{x}\rrbracket_{PPFA}^{FA}}) \supseteq D^c([\llbracket{\sf C}\rrbracket_{PPFA}^{FA}]_{\llbracket{x}\rrbracket_{PPFA}^{FA}})$ \hspace{\fill}from \tref{congpar}

$ \forall_{x \in Det_{PPFA}}. D^c([\llbracket{\sf A}\rrbracket_{PPFA}^{FA}]_x) \supseteq D^c([\llbracket{\sf C}\rrbracket_{PPFA}^{FA}]_x) $\hspace{\fill} \lref{suf} part 1(a)

$\llbracket{\sf A}\rrbracket_{PPFA}^{FA}\sqsubseteq_{PPFA}  \llbracket{\sf C}\rrbracket_{PPFA}^{FA}$\hspace{\fill} from \dref{pref}

\noindent 4.  
Re: show $vA^{FA}_{PPFA}$ is monotone:   ${\sf A}\sqsubseteq_{PPFA}  {\sf C}\Rightarrow vA^{FA}_{PPFA}{\sf A}\sqsubseteq_{FA}  vA^{FA}_{PPFA}{\sf C}$

From ${\sf A}\sqsubseteq_{PPFA}  {\sf C}$ and definitions we have:
$\forall_{x\in\Xi_{PPFA}}. D^c([{\sf  A}]_x) \supseteq D^c([{\sf  C}]_x) $ 

as $Det_{PPFA} \subset \Xi_{PPFA}$ we have 

$\forall_{x\in Det_{PPFA}} .D^c([{\sf  A}]_x) \supseteq D^c([{\sf  C}]_x) $  \hspace{\fill}(2)


For all $o$ in $Tr^c(vA^{FA}_{PPFA}([{\sf  C}]_x))$ there must exist a $d$ in $D^c([{\sf  C}]_x) $ such that $o\in dom(d)$  and from (2) we know that $d$ is in $D^c([{\sf  A}]_x) $  and with $o\in dom(d)$  we can conclude that $o$ in $Tr^c(vA^{FA}_{PPFA}([{\sf  A}]_x))$ so: 

$\forall_{x\in Det_{PPFA}}. Tr^c(vA^{FA}_{PPFA}([{\sf  A}]_x)) \supseteq Tr^c(vA^{FA}_{PPFA}([{\sf  C}]_x))$

$\forall_{x\in Det_{PPFA}}. Tr^c([vA^{FA}_{PPFA}{\sf  A}]_{vA^{FA}_{PPFA}x}) \supseteq Tr^c([vA^{FA}_{PPFA}{\sf  C}]_{vA^{FA}_{PPFA}x})$\hspace{\fill}\tref{congpar} 

$\forall_{x\in Det_{FA}} .Tr^c([vA^{FA}_{PPFA}{\sf  A}]_{x}) \supseteq Tr^c([vA^{FA}_{PPFA}{\sf  C}]_{x})$   \hspace{\fill}from \lref{suf} part 1(b)

$\forall_{x\in \Xi_{FA}} .Tr^c([vA^{FA}_{PPFA}{\sf  A}]_{x}) \supseteq Tr^c([vA^{FA}_{PPFA}{\sf  C}]_{x})$   \hspace{\fill}from \lref{suf} part 3

$vA^{FA}_{PPFA}{\sf A}\sqsubseteq_{FA}  vA^{FA}_{PPFA}{\sf C}$ \hspace{\fill} \dref{ref}

\lend
 
 The embedding $\llbracket\_\rrbracket_{PPFA}^{FA}$ can be used to add probability to a non-probabilistic finite automata during the stepwise development, \ie\ refinement, of a model or specification. This use of Galois connections is nothing new but to the best of our knowledge it is the first time it has been used to allow the introduction of probability part of the way through the development of a process. 
 
 \section{Conclusions}
 
Others have used the same testing framework to treat probabilistic processes, but in one notable case \cite{DGHMZ07} it was found that many of the expected algebraic results were false according to the testing used. This meant the abandonment of testing as a basis for refinement and a notion of simulation was introduced. We believe that the reason that many of the ``sanity checks" turned out to be false for the testing-based refinement in that paper was that the original formalisation of nondeterminism found in non-probabilistic systems was kept and that this led to problems when probabilistic tests on nondeterministic probabilistic systems were considered. 

Instead of abandoning refinement based on testing, we handle nondeterminism in a way that is compatible with probability, rather than using the original formalisations of testing nondeterminism found in non-probabilistic systems. 

We also note that, having shown we can (as a (vertical) refinement) move from non-probabilistic models to probabilistic ones (and back again, if we wish), the introduction of probabilities can happen as a design step \emph{during} development of a system via refinement steps. So, we are free to take a very general non-probabilistic specification and, if it turns out to be necessary to do so to deal with some aspects of the specification, introduce probabilities as we make progress towards a more concrete form of the system. We have not yet explored this possibility, but it does introduce another freedom to the developer which might turn out to be useful.

The framework we have introduced in this paper is really only a first step towards a sensible language for specifying systems containing probability. What still needs to be done is to recognise that some sorts of probabilistic choice do not ``make sense", \ie\ that there are right and wrong places to use such choice. For example, if we have a vending machine with two buttons on, one for tea and one for coffee, it clearly does not make sense to specify the choice here as a probabilistic one---the vending machine would be a very odd one if it allowed me to choose tea only 75\% of the time!

On the other hand, it does make sense (though perhaps inventing plausible uses for such a thing might be hard!) to specify a robot which can make choices from a vending machine that offers tea or coffee, where the robot prefers tea over coffee, so it chooses tea 75\% of the time. 

The difference between these two cases is one of \emph{causality}. The robot's actions cause the vending machine's, and not \emph{vice versa}. So, our specification language would need to allow us to make this distinction and, most helpfully, only allow probabilistic choice to be specified in situations where it makes sense, as in the case of specifying the robot. We have done previous work on adding causality (back) into process algebras, and the work presented here forms the basis for a probabilistic causal process algebra (CPA) \cite{ReS05d}, or for a probabilistic language for interactive branching processes (IBPs) \cite {ReS10} which we have also talked about before, which forms the subject of another paper yet to be published.

A final interesting point to note is that, because we can always migrate probabilities on actions right up the probabilities on start states, we have a normal form for our automata. In this form, the only place that probabilities appear is on the start states (so the only non-trivial probabilistic distribution over states is the start-state distribution). This makes it very clear that one needs only one roll (of dice with enough faces) in order to conduct a probabilistic computation.
                                                                                                                                                                                                                        
\bibliographystyle{eptcs}
\bibliography{refs} 

\end{document}